\documentclass[aps,prx,amsfonts,floatfix,superscriptaddress,longbibliography,twocolumn,footinbib]{revtex4-2}
\usepackage[utf8]{inputenc}
\usepackage{amsmath}    
\usepackage{amssymb}
\usepackage{mathbbol}
\usepackage{graphicx}   
\usepackage{verbatim}   
\usepackage{subfigure}  
\usepackage{hyperref} 
\usepackage{scalerel}
\usepackage{cancel}
\usepackage{soul}
\usepackage[normalem]{ulem}
\usepackage[dvipsnames]{xcolor}
\usepackage{siunitx}
\usepackage{physics}
\usepackage{enumerate} 
\usepackage{empheq}

\begin{document}

\title{Phase transitions, symmetries, and tunneling in Kerr parametric oscillators}

\author{Miguel A. Prado Reynoso} 
\affiliation{Nonlinear Dynamics, Chaos and Complex Systems Group, Departamento de F\'{i}sica,  Universidad Rey Juan Carlos, Tulip\'{a}n s/n, 28933 M\'{o}stoles, Madrid, Spain}
\affiliation{Instituto de Ciencias Nucleares, Universidad Nacional Aut\'onoma de M\'exico, 04510 Cd. Mx., Mexico}
\affiliation{Department of Physics, University of Connecticut, Storrs, Connecticut 06269, USA}

\author{Edson M. Signor}
\affiliation{Department of Physics, University of Connecticut, Storrs, Connecticut 06269, USA}

\author{Jamil Khalouf-Rivera}
\affiliation{Departamento de Ciencias Integradas y Centro de Estudios Avanzados en F\'isica, Matem\'aticas y Computaci\'on, Universidad de Huelva, Huelva 21071, Spain}

\author{Alexandre D. Ribeiro}
\affiliation{Department of Physics, University of Connecticut, Storrs, Connecticut 06269, USA}
\affiliation{Departamento de F\'{\i}sica, Universidade Federal do Paran\'a, 81531-990, Curitiba, PR, Brazil}

\author{Francisco P\'erez-Bernal}
\affiliation{Departamento de Ciencias Integradas y Centro de Estudios Avanzados en F\'isica, Matem\'aticas y Computaci\'on, Universidad de Huelva, Huelva 21071, Spain}

\author{Lea F. Santos}
\affiliation{Department of Physics, University of Connecticut, Storrs, Connecticut 06269, USA}

\begin{abstract}
Quantum Kerr parametric oscillators (KPOs) are systems out of equilibrium with a wide range of applications in quantum computing, quantum sensing, and fundamental research. They have been realized in superconducting circuits and photonic platforms. In this work, we explore the onset of ground-state and excited-state quantum phase transitions in KPOs, focusing on the role of the phase-space rotational symmetry when the driving frequency is $\mu$ times the oscillator’s natural frequency, specifically for $\mu=1,2,3,4$. These cases are experimentally accessible in superconducting circuits, where the Floquet quasienergy spectrum can also be studied as a function of tunable control parameters. Using the classical Hamiltonian of the system, we identify the critical points associated with quantum phase transitions and analyze the emergence of both real and avoided level crossings, examining their influence on the energy spectrum and tunneling dynamics. Our findings provide insights into the engineering of robust quantum states,  quantum dynamics control, and  onset of quantum phase transitions with implications for critical quantum sensing.
\end{abstract}

\maketitle

\section{Introduction}
\label{Intro}

A parametric oscillator is  an oscillator with one or more time-dependent parameters, resulting in the amplification or modification of its oscillations. In this work, we focus on quantum  oscillators with Kerr nonlinearity, where parametric oscillations arise from an external drive. These Kerr parametric oscillators (KPOs) are versatile platforms for various quantum applications, ranging from the generation of squeezed states~\cite{Schnabel2017} to studies of tunneling and quantum activation~\cite{Marthaler2006,Marthaler2007,DykmanBook2012}. They can be realized in multiple experimental setups, including trapped ions~\cite{Ding2017}, optical devices~\cite{Roy2021,Perez2023}, and superconducting circuits~\cite{Kirchmair2013,Mirrahimi2014,Wang2019KPO,Frattini2024},  and potentially in semiconductor microcavities with exciton-polaritons~\cite{Ghosh2020, Byrnes2014} and nanomechanical resonators~\cite{Marti2024,Woolley2008}. 

Superconducting KPOs have been particularly prominent in quantum computing~\cite{Goto2016,Goto2019}, where they are extensively studied for the generation of Schr\"odinger cat states~\cite{Puri2017} used in Kerr-cat qubits~\cite{Grimm2020}, quantum error correction schemes~\cite{Puri2019,Kwon2022} for fault-tolerant computation, 
Ising machines~\cite{Kanao2021,Yamaji2023}, where KPO networks solve combinatorial optimization problems, and for addressing complex phenomena in chemistry~\cite{Cabral2024,Dutta2024,Albornoz2024}. Recently, they have also been exploited for
critical quantum sensing~\cite{Candia2023,Beaulieu2024,Chavez2024} where phase transitions are used to improve sensitivity.

Beyond technological applications, KPOs serve as quantum simulators of excited-state quantum phase transitions (ESQPTs)\cite{Chavez2023,Garcia-Mata2024a} (also known as ``spectral kissing''~\cite{Frattini2024}) with implications for quantum dynamics in Hilbert space and phase space~\cite{Chavez2023}. ESQPTs are generalizations of ground state quantum phase transitions (QPTs) to  excited states with consequences to the spectrum~\cite{Cejnar2008,Caprio2008,Cejnar2021}, structure of the eigenstates~\cite{SantosBernal2015,Bernal2017,Santos2016}, and dynamics of the quantum system~\cite{Santos2016,Lobez2016,Kloc2018,Pilatowsky2020,Chavez2023,Prado2023}. Their critical points can be reached either by varying one or several control parameter for a given energy level or by fixing the control 
parameters and increasing the excitation energy. Although QPTs and ESQPTs are strictly realized only in the large system size limit, precursors of their existence can be observed for finite system sizes~\cite{Iachello2004}.
KPOs have also been used to explore dissipative phase transitions~\cite{Beaulieu2023} and quantum chaos~\cite{Burgelman2022,Cohen2023,Chavez2025, Goto2021}.

For experimentally relevant Kerr nonlinearities and drive amplitudes~\cite{Frattini2024}, KPOs within the rotating-wave approximation can be described by effective time-independent Hamiltonians~\cite{DykmanBook2012,Venkatraman2022,xiao2023diagrammatic}. These Hamiltonians exhibit  symmetries~\cite{Ruiz2023,Iachello2023} that influence their spectral structure, phase diagrams, and dynamical properties. Given the importance of quantum phase transitions and symmetries in KPOs for critical sensing, dynamical control, and quantum state stabilization, we carry out a systematic investigation of QPTs, ESQPTs, and quantum tunneling in four different KPOs' effective time-independent Hamiltonians. 

We consider systems where the driving frequency is close to $\mu$ times the oscillator's natural frequency, with $\mu = 1, 2, 3, 4$ resulting in four distinct effective Hamiltonians denoted by $\hat{H}_{\mu}$.  The QPTs are classified using Ehrenfest’s criterion, and the classical limit of the system is analyzed to identify the critical points associated with both QPTs and ESQPTs. For each effective Hamiltonian $\hat{H}_{\mu}$, we construct its phase diagram  and analyze the role of spontaneous symmetry-breaking in the emergence of the various phases. 

The analysis of phase transitions does not go beyond four-photon drive, because the resulting Hamiltonian for $\mu>4$ is unbounded~\footnote{The study of driven nonlinear oscillators with $\mu>4$, including translational symmetry, has been done in the context of phase space crystals~\cite{Guo2013, Liang2018, PhSpBook}.}. In fact, we discuss that even for $\mu=4$, there is a range of parameters for which the Hamiltonian is unbounded. As we approach this point, the quantum system undergoes changes in the ground state and excited states that cannot be explained within the classical limit used here.

Furthermore, by computing the energy spectrum of $\hat{H}_{\mu}$ as a function of the control parameters, we identify real and avoided level crossings, which are directly linked to quantum tunneling phenomena.  Quantum tunneling allows access to classically unconnected regions of phase space. The ability to enhance or suppress tunneling through knowledge of the spectrum provides a tool for controlling quantum dynamics. 

The KPO effective Hamiltonians, $\hat{H}_\mu$, that we investigate exhibit $\mathbb{Z}_{\mu}$ symmetry, being invariant under discrete rotations by $2 \pi/\mu$ in phase space. The Hilbert space correspondingly splits into $\mu$ symmetry sectors, leading to real and avoided level crossings that affect quantum tunneling~\cite{Prado2023,Venkatraman2024,Iyama2024}. While avoided crossings facilitate tunneling by enabling spread within the same symmetry sector, real crossings between different symmetry sectors can suppress tunneling, effectively protecting states from hybridization. Our results offer a systematic framework for controlling tunneling dynamics through spectral engineering. These findings advance the understanding of symmetry-induced phenomena in nonlinear quantum oscillators and have direct implications for experimental platforms that combine Kerr nonlinearities and external drives.

The work is divided as follows. Section~\ref{Model} introduces the quantum models for the KPOs along with their corresponding classical Hamiltonians, emphasizing the role of symmetries in shaping the phase-space structure. In Sec.~\ref{CQ_PT}, we analyze the QPTs and ESQPTs for one- to four-photon drives, highlighting the connections between the quantum phase transitions and the critical points of the system's classical counterparts. We also present an exception to this picture, where the changes to the quantum system are not captured by the mean-field limit. Section ~\ref{Q_tunnel} investigates the effects of symmetries on the tunneling processes of KPOs. Lastly, in Sec.~\ref{Conclusion}, we summarize our key findings and discuss their potential applications in quantum technologies.

\section{Model and symmetries}
\label{Model}

This section describes the quantum model and its corresponding classical Hamiltonian, and identifies the symmetries of the system.

\subsection{Quantum Model}

We consider a Kerr nonlinear oscillator driven at a frequency $\omega_d$ that is close to $\mu$ times the oscillator's natural frequency, $\omega_0$, with $\mu=1,2,3,4$. In the frame rotating with frequency $\omega_d/\mu$, the effective time-independent Hamiltonian is given by
\begin{equation}
\frac{\hat{H}_\mu}{\hbar K} = \hat{H}_{\delta} -
\xi_\mu \left(\hat{a}^{\dagger \mu} +  \hat{a}^\mu\right),    
\label{Eqeff} 
\end{equation}
where we fix $\hbar=1$,
\begin{equation}
\hat{H}_{\delta} \equiv -\delta \, \hat{a}^{\dagger} \hat{a} + \hat{a}^{\dagger 2} \hat{a}^2 ,
\label{Eq:Hqdelta0}
\end{equation}
$\delta \equiv (\omega_0 - \omega_d/\mu )/K$ denotes the frequency detuning, and
$\xi_{\mu}\equiv \epsilon_{\mu}/K$ is the ratio between the drive amplitude, $\epsilon_{\mu}$, and the Kerr amplitude, $K$. Our control parameters are $\delta$ and $\xi_{\mu}$. 

The KPOs Hilbert space is of infinite dimension, hence we truncate it at a given size $N$, that ensures the convergence of the energy levels under consideration. Throughout this work, energy is denoted by $E$, which is written in units of $K$.

In superconducting circuit experiments, the Hamiltonian's sign is the inverse of that in Eq.~(\ref{Eqeff}). Our choice is made for convenience, implying that the bottom of a metapotential in this paper corresponds to the peak of an inverted metapotential 
in experimental setups, where dissipation brings the system toward the attractor.

\subsection{Symmetries}

The Hamiltonian $\hat{H}_\mu$ in Eq.~(\ref{Eqeff}) commutes with the symmetry operator
\begin{equation}
 \hat{\mkern-3mu \cal \scriptstyle S}_{\mu} 
\equiv \mbox{e}^{-i 2\pi \hat{n}/\mu},  
\label{Eq:SymOp}
\end{equation}
where $\hat{n}\equiv \hat{a}^\dagger \hat{a}$ is the number operator. The Fock states $|n\rangle$ are eigenstates of $\hat{\mkern-3mu \cal \scriptstyle S}_{\mu}$ with only $\mu$ distinct eigenvalues. The symmetry partitions the Hilbert space into $\mu$ invariant sectors, each associated with a distinct eigenvalue of the symmetry operator. The symmetry associated with $\hat{\mkern-3mu \cal \scriptstyle S}_{\mu}$ is $\mathbb{Z}_{\mu}$, meaning that the Hamiltonian is invariant under discrete rotations by $2\pi/\mu$ in phase space.

The emergence of symmetry sectors is important for understanding the properties of the energy spectrum as a function of the Hamiltonian parameters. According to the Wigner–von Neumann theorem~\cite{vonNeuman1929,Demkov2007}, real degenerate energy eigenvalues can only happen for eigenstates belonging to different symmetry sectors, whereas quasi-degenerate energy levels, associated with avoided crossings, occur within the same symmetry sector. The effects of level crossings on  quantum tunneling are analyzed in Sec.~\ref{Q_tunnel}.

\subsection{Classical Hamiltonian}
\label{ClassLim}

By analyzing the classical limit of the system, we determine the critical points, assess their stability, and calculate their associated energies, enabling the construction of phase diagrams. The classical counterpart of Eq.~\eqref{Eqeff}, derived in App.~\ref{Ap:CH}, is 
\begin{equation}
\frac{{H}_\mu^c}{ K} = 
{H}_\delta^{\mathrm{c}} - \frac{2\xi_\mu}{2^{\mu/2}} \mathcal{F}_\mu , 
\label{EqHcl}
\end{equation}
where
\begin{eqnarray}
&{H}_\delta^c& \equiv -\frac{\delta}{2} \left(q^2 + p^2\right)+ \frac{1}{4} \left(q^2 + p^2\right)^2 , 
\label{Eq:Hdeltac}
\\
&\mathcal{F}_\mu& \equiv \mathrm{Re}[(q+ip)^\mu].
\end{eqnarray}
The $\mathcal{F}_\mu$ function stems from the $\mu$-photon drive term and
\begin{equation}
\begin{aligned}
\mathcal{F}_1 &= q,\\ 
\mathcal{F}_2 &= q^2-p^2,\\ 
\mathcal{F}_3 &= q^3-3qp^2,\\
\mathcal{F}_4 &= q^4  - 6q^2p^2 + p^4.
\end{aligned}
\nonumber
\end{equation}

The quantum symmetry associated with the operator in Eq.~\eqref{Eq:SymOp} is classically manifested as a ${\cal C}_\mu$ rotational symmetry around the phase-space origin. To show this, consider the complex variable $z=|z|\mbox{e}^{i\theta}= (q+ip)/\sqrt2$ and its complex conjugate $z^*$. In terms of these new variables, the classical Hamiltonian in Eq.~\eqref{EqHcl} becomes ${H}_\mu^z  = - \delta |z|^2  +  |z|^4  - {\xi}_\mu \left( z^{\mu} + z^{*\mu} \right)$. Rotating the phase-space energy surface by an angle $\beta$ around the origin corresponds to changing the variable $z$ as $z \to z\,\mbox{e}^{-i\beta}$, which keeps the classical Hamiltonian invariant for $\beta =(2\pi/\mu)$.

\section{Classical and Quantum phase transitions}
\label{CQ_PT}

In classical phase transitions, the stationary points of the system's Hamiltonian are used for identifying critical points associated with changes in the system's behavior. As the control parameters are varied, changes in the number, type, and stability of the stationary points indicate phase transitions. 

The stationary points $(q_s,p_s)$ of the classical Hamiltonian in Eq.~\eqref{EqHcl}, also referred to as critical points,  are obtained imposing equilibrium conditions within Hamilton's equations of motion, 
\begin{equation}
\begin{split}
-p_s \left[
\delta -  \left(q_s^2 + p_s^2 \right)\right] - \left. \frac{2\xi_\mu}{2^{\mu/2}}
\frac{\partial  \mathcal{F}_\mu}{\partial p} \right|_{q_s,p_s}
&=0,
\\ 
q_s \left[
\delta - \left(q_s^2 + p_s^2 \right)\right] + \left. \frac{2\xi_\mu}{2^{\mu/2}}
\frac{\partial  \mathcal{F}_\mu}{\partial q} \right|_{q_s,p_s}
&=0 .
\end{split}
\label{Hamil_eq}
\end{equation}
The analysis of these points provides insights into QPTs and ESQPTs, which emphasizes
the connection between the classical and quantum descriptions of the system.  QPTs manifest as  abrupt changes in the ground state of a system for particular values of the control parameters (critical control parameter values)~\cite{Iachello2004}, while ESQPTs are associated with degenerate or quasi-degenerate excited eigenstates, and with divergences or discontinuities in the density of states (DOS) or its derivatives~\cite{Stransky2014,Stransky2016,Cejnar2021}. 

In the following subsections, from Sec.~\ref{Sec:One} to Sec.~\ref{Sec:Four}, we present a systematic analysis of the phase space and phase diagram for each classical Hamiltonian $H^c_{\mu}$, exploring their connection with QPTs and ESQPTs. Before doing that, we describe below in  Sec.~\ref{Sec:NoDrive} the non-driven case ($\xi_{\mu}=0$). 

The phase diagrams for all Hamiltonians $H^c_{\mu}$ are symmetric under the sign change of $\xi_\mu$. For this reason, it is sufficient to analyze the phase transitions for $\xi_\mu > 0$, noting that for $\xi_\mu < 0$ the critical values $p_s$ and $q_s$ are rotated by an angle $\pi/\mu$. We label the phases for $\xi_\mu > 0$ with roman numerals (I, II, ...) and those for $\xi_\mu < 0$ are denoted with a tilde ($\widetilde{\mathrm{I}}$, $\widetilde{\mathrm{II}}$,...). In all phase diagrams presented in this work, solid lines indicate boundaries between two regions undergoing a QPT, where the separated regions may also exhibit different ESQPTs, while dotted lines separate regions that do not undergo a QPT but differ in the number of ESQPTs they exhibit.

\subsubsection{Phase-transition order} 

In App.~\ref{Ap:Op}, we explain how to determine the order of the phase transition for KPOs using Ehrenfest's criterion. This is done by computing the lowest energy of the system and studying its dependence on the control parameters $\delta$ and $\xi_\mu$.

\subsubsection{Density of states and ESQPT} 

The classical DOS is defined as the available phase-space volume at a given energy $E$, while the quantum DOS is obtained by counting the number of energy levels in a small energy window centered at $E$. At large quantum numbers, the two densities converge~\cite{GutzwillerBook}. The comparison of  classical and quantum distributions provides a means for tracking the classical origins of the quantum transitions.

In systems with one degree of freedom, an ESQPT is characterized by a logarithmic peak (denoted here by ESQPT$_{\text{peak}}$) or by a discontinuous step (referred to as ESQPT$_{\text{step}}$) in the quantum DOS~\cite{Stransky2016,Cejnar2021}. These features in the vicinity of an ESQPT critical energy reflect  changes in the topology of the classical phase space. A divergent peak in the DOS is associated with the emergence of a hyperbolic (saddle) point in phase space and the discontinuous step with the presence of a local maximum or a local minimum.

\subsubsection{Non-driven KPO: $\xi_{\mu}=0$} 
\label{Sec:NoDrive}

\begin{figure}[h]
\begin{center}
\includegraphics[scale=0.80]{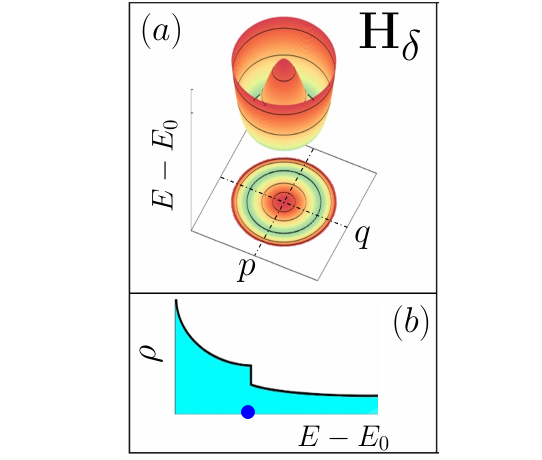}
\end{center}
\caption{Non-driven KPO. Panel (a) illustrates the energy landscape (top) of the classical Hamiltonian $H_{\delta}^c$ in Eq.~(\ref{Eq:Hdeltac}) with respect to the lowest energy $E_0$ and the projection of the landscape in the two-dimensional phase space (bottom). Panel (b) represents the quantum DOS. The blue circle indicates the energy of the local maximum, where the DOS presents a discontinuous step.}
\label{fig:nodrive}
\end{figure}

In the special case of $\xi_\mu = 0$, the classical energies, according to Eq.~(\ref{Eq:Hdeltac}), are  $E^c= (q^2 + p^2)[(q^2 + p^2)/2 - \delta]/2$, and the quantum levels, according to Eq.~(\ref{Eq:Hqdelta0}), are $E^q_n = n^2 - n(1+\delta)$, where $n=0, 1, 2, \ldots$.
The changes in the energies occur along the line of $\delta$ values. When $\delta \leq 0$, the classical Hamiltonian exhibits a single global minimum with energy $E^c_0=0$, while in the quantum spectrum, the ground-state energy $E^q_0=0$ is non-degenerate for $\delta<0$ and twofold degenerate at $\delta=0$. When $\delta>0$, there is a change in the stationary points that results in a second-order phase transition. The classical system now presents a circular set of global minima satisfying the equation $q_s^2 + p_s^2 = \delta$ with $E^c_0=-\frac14\delta^2$.  This set of global minima corresponds to the green circle in the phase space of Fig.~\ref{fig:nodrive}(a). The surface on top of Fig.~\ref{fig:nodrive}(a) illustrates the energy landscape and its projection in the two-dimensional phase space is shown in the bottom of the panel. The ground-state energy is twofold degenerate and given by $E^q_0=-\frac{\delta}{2} (1 + \frac{\delta}{2})$ whenever $\delta$ is even and it loses the degeneracy for other  values of $\delta$, when $E^q_0 = -(\lfloor \frac{\delta}{2}\rfloor + 1) (\delta - \lfloor \frac{\delta}{2}\rfloor)$. 

Figure~\ref{fig:nodrive}(b) illustrates the quantum DOS for $\delta >0$. It exhibits a step discontinuity at the energy indicated with a blue circle. This is the critical energy of an ESQPT$_{\text{step}}$, which converges to the energy of the local maximum in the classical limit. Below the ESQPT$_{\text{step}}$ energy, there can be pairs of degenerate quantum levels associated with pairs of same-energy classical trajectories, where one trajectory is in an inner circle around the global maximum and the other is in an outer circle. Above the ESQPT$_{\text{step}}$ critical energy, the degeneracies are lifted.

\subsection{One-photon drive}
\label{Sec:One}

\begin{figure*}[t]
\begin{center}
\includegraphics[scale=0.9]{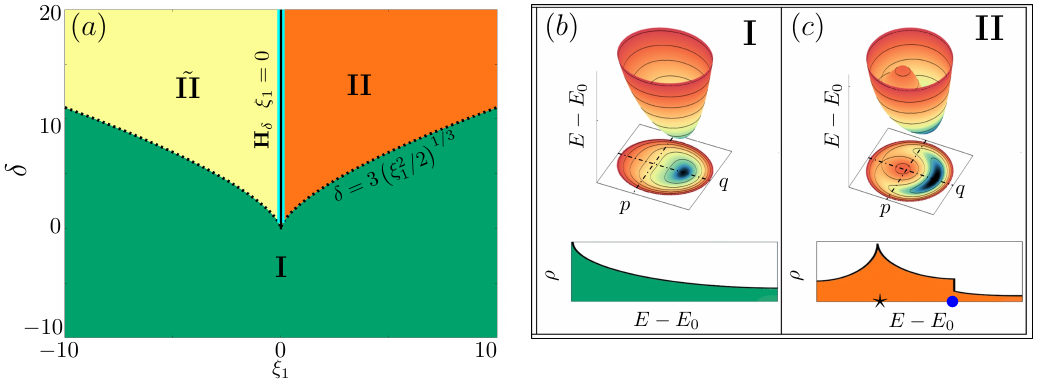}
\end{center}
\vspace{-0.5cm}
\caption{One-photon KPO. Panel (a) shows the phase diagram for the detuning parameter $\delta$ and the one-photon drive amplitude $\xi_1$. The QPT is represented by a solid line, while the dotted lines indicate ESQPTs. Panels (b) and (c) illustrate the energy landscape and phase-space structure (top) and the DOS (bottom) for  region I and region II, respectively.}
\label{fig:clmu1}
\end{figure*}

The KPO in the presence of coherent (one-photon) driving is the simplest experimental case. From a theoretical point of view, its Hilbert space is not divided into symmetry sectors, since the operator $\hat{\mkern-3mu \cal \scriptstyle S}_1$ in Eq.~(\ref{Eq:SymOp}) is simply the identity, so any energy level crossing that may occur in this case is an avoided crossing~\cite{Chavez2024}.

Solving the system of equations in Eq.~\eqref{Hamil_eq} for the classical Hamiltonian with $\mu=1$,
\begin{equation}
\begin{split}
{H}_1^c &= -\frac{\delta}{2} \left(q^2 + p^2\right)+ \frac{1}{4} \left(q^2 + p^2\right)^2 - \sqrt{2}\xi_1 q,
\end{split}
\label{qH1}
\end{equation}
shows that all stationary points for $\xi_1 \neq 0$ have zero momentum.  Finding the values of position of these points requires solving the cubic equation $q_s^3 - \delta \, q_s - \sqrt{2}\xi_1= 0$, which gives the following critical points $\bold{r}=\{ q_s, p_s\}$ and their corresponding classical energies ${\cal E}$,
\begin{equation}
\begin{array}{ll}      
    \bold{r}_{1} = \left\{ \eta_+, 0 \right\} ,
    & \hspace{0.3 cm} {\cal E}_{\bold{r}_1} = H_1^{\mathrm{c}}(q_{s_1},0)  ,
     \\   
    \bold{r}_{2} = \left\{ -\dfrac12 \eta_+  + \dfrac{i \sqrt{3}}{2} \eta_-   , 0 \right\} ,
    & \hspace{0.3 cm} {\cal E}_{\bold{r}_2} = H_1^{\mathrm{c}}(q_{s_2},0)  ,
     \\  
    \bold{r}_{3} = \left\{ -\dfrac12 \eta_+  - \dfrac{i \sqrt{3}}{2} \eta_-, 0 \right\} ,
    & \hspace{0.3 cm} {\cal E}_{\bold{r}_3} = H_1^{\mathrm{c}}(q_{s_3},0)  ,
\end{array}
\label{eq:fxp_1}
\end{equation}
where $\eta_\pm = \beta_+ \pm \beta_-$ with $\beta_{\pm} = \big[\xi_1/\sqrt{2} \pm \sqrt{\Delta_{\mathrm{dsc}}} \big]^{1/3}$ and the 
discriminant, 
\begin{equation}
\Delta_{\mathrm{dsc}}\equiv
\frac{\xi_1^2}{2} - \frac{\delta^3}{27},
\nonumber
\end{equation}
determines the number of real solutions of the cubic equation. 
For $\Delta_{\mathrm{dsc}} > 0$, the equation has only one real solution.  Along the boundary line $\delta=3(\xi_1^2/2)^{1/3}$, two solutions exist. 
When $\Delta_{\mathrm{dsc}}<0$, the equation admits three distinct real solutions.  

The phase diagram in Fig.~\ref{fig:clmu1}(a) is constructed using the discriminant, with each color representing a distinct phase. A solid line marks the boundary between two regions undergoing a QPT. These regions may also exhibit different ESQPTs. A dotted line separates regions that do not experience a QPT, but differ in the number of ESQPTs.

In region I of Fig.~\ref{fig:clmu1}(a), where $\delta < 3 (\xi_1^2/2 )^{1/3}$, there is only one stationary point. This is a stable global minimum that approaches the phase-space origin as $\xi_1$ approaches zero. The energy landscape is characterized by a single well, as seen in the top of Fig.~\ref{fig:clmu1}(b). The bottom of this figure shows the quantum DOS, which decreases monotonically with increasing energy.

In region II of Fig.~\ref{fig:clmu1}(a), where $\delta > 3(\xi^2_1/2)^{1/3}$, there are three critical points: a global minimum, a hyperbolic point, and a local maximum, as seen in the energy landscape and phase space in Fig.~\ref{fig:clmu1}(c). The energy contour crossing at the hyperbolic point is the separatrix that divides the phase space into different dynamical regions. In the energy interval above the separatrix and below the local maximum, the spectrum can exhibit avoided crossings associated with same-energy trajectories, one trajectory close to the local maximum and the other in the phase-space region that encompasses all critical points~\cite{Chavez2024}.

The DOS at the bottom of Fig.~\ref{fig:clmu1}(c) shows a peak at the energy indicated with a star and this peak diverges logarithmically as we approach the classical limit. This accumulation of energy levels in the DOS signals an  ESQPT$_{\text{peak}}$. The separatrix energy is close to the ESQPT$_{\text{peak}}$ critical energy and coincides with it in the classical limit. At an even larger energy, one reaches the ESQPT$_{\text{step}}$, characterized by the step discontinuity in the DOS. This happens at the energy marked with a blue circle, which coincides with the energy of the local maximum in the classical limit. 

As the system transitions from region I to region II, it undergoes significant structural changes in the energy landscape and phase-space topology, which are reflected in the DOS. While region I lacks any ESQPTs, region II features two distinct ESQPTs. At the transition line [dotted line at $\delta=3(\xi_1^2/2)^{1/3}$], the system exhibits one global minimum and an inflection point, which corresponds to a degenerate critical point, where the DOS diverges logarithmically.

To move from region II to region $\widetilde{\mathrm{II}}$, one needs to pass through the line of $\xi_{1}=0$, already described in Fig.~\ref{fig:nodrive} and indicated in Fig.~\ref{fig:clmu1}(a) with a solid  cyan line. This corresponds to a first-order QPT, where the minimum and hyperbolic critical points of region II merge into a circle of global minima when $\xi_{1}=0$, before reappearing in region $\widetilde{\mathrm{II}}$. This is a particular kind of first-order QPT, without coexistence of global minima. A similar first-order transition happens in nuclear shapes~\cite{Jolie2001, Cejnar2009, Cejnar2010} and in the Lipkin-Meshkov-Glick model~\cite{Romera2014}.

\subsection{Two-photon drive}

\begin{figure*}[!t]
\begin{center}
\includegraphics[scale=0.9]{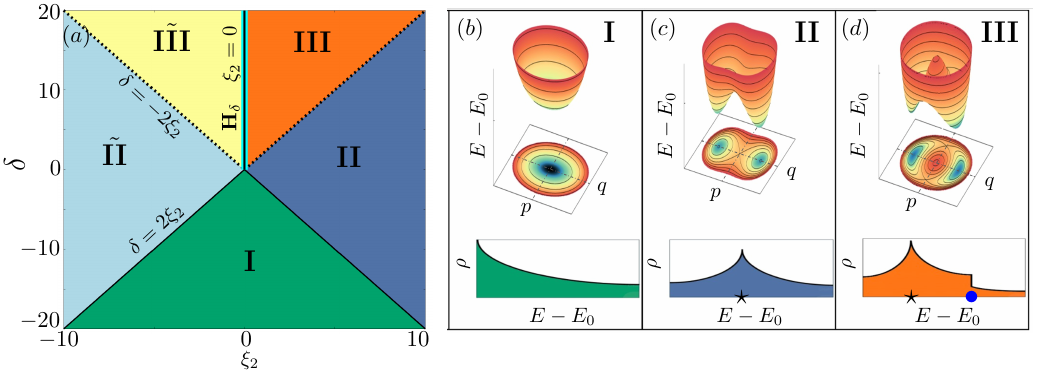}
\end{center}
\vspace{-0.5cm}
\caption{Two-photon KPO. Panel (a) shows the phase diagram for the detuning parameter $\delta$ and the two-photon drive amplitude $\xi_2$. QPTs (ESQPTs) are represented by solid (dotted) lines. Panels (b)-(d) illustrate the energy landscape and phase-space structure (top) and the DOS (bottom) for three regions appearing in the phase diagram: region I (b), II (c), and III (d).}
\label{fig:clmu2}
\end{figure*}

The two-photon KPO ($\mu=2$) was analyzed in detail in~\cite{Prado2023} after its experimental realization in~\cite{Venkatraman2024}.
The classical Hamiltonian is given by
\begin{equation}
\begin{split}
{H}_2^{\mathrm{c}} &= -
\frac{\delta}{2} \left(q^2 + p^2\right)+ \frac{1}{4} \left(q^2 + p^2\right)^2 - \xi_2\left( q^2 - p^2 \right).
\end{split}
\label{qH2}
\end{equation}
Solving the system of equations in Eq.~\eqref{Hamil_eq}, we distinguish five critical points $\bold{r}=\{q_s, p_s\}$, which are listed below together with their corresponding energies ${\cal E}$,
\begin{equation}
\begin{array}{ll}
\bold{r}_0 = \{0,0\},
& \hspace{0.3 cm} {\cal E}_{\bold{r}_0} = 0 ,
\\
\bold{r}_1^{\pm} = \Big\{0, \pm\sqrt{\delta-2\xi_2}\Big\}, 
& \hspace{0.3 cm}  {\cal E}_{\bold{r}_1^{\pm}} = -\left(\dfrac{\delta-2\xi_2}{2}\right)^2 ,
\\ 
\bold{r}_2^{\pm} = \Big\{\pm\sqrt{\delta+2\xi_2},0 \Big\},
& \hspace{0.3 cm}  {\cal E}_{\bold{r}_2^{\pm}} = -\left(\dfrac{\delta+2\xi_2}{2}\right)^2.
\end{array}
\label{eq:fxp_2}
\end{equation}

The phase-space origin is a stationary point for any value of $\delta$ and $\xi_2$, while the critical points $\bold{r}_1^{\pm}$ ($\bold{r}_2^{\pm}$) exist only  if $\delta>2\xi_2$ ($\delta>-2\xi_2$).  According to the stability of these points, three main scenarios are identified in the phase diagram of Fig.~\ref{fig:clmu2}(a), as explained next. 

In region I of  Fig.~\ref{fig:clmu2}(a), where $\delta \leq - 2|\xi_2|$, there is only one stationary point at $\bold{r}_0$, which is a global minimum, as seen at the top of Fig.~\ref{fig:clmu2}(b). Accordingly, the density of states at the bottom of Fig.~\ref{fig:clmu2}(b) decreases monotonically as the energy increases.

In region II of Fig.~\ref{fig:clmu2}(a), where $\xi_2>0$ and $-2\xi_2< \delta \leq 2\xi_2$, we have three stationary points. Two of these points are stable global minima at $\bold{r}_2^{\pm}$ and $\bold{r}_0$ becomes now an unstable hyperbolic point. The DOS shown at the bottom of Fig.~\ref{fig:clmu2}(c) peaks at the energy of the hyperbolic point, as typical of an ESQPT$_{\text{peak}}$. 

As the system transitions from region I to region II, it undergoes a second-order QPT, where the single minimum at the origin bifurcates into two new minima at $\bold{r}_2^\pm$. The order of this QPT is established using Ehrenfest's criterion, as shown in App.~\ref{Ap:Op}. Furthermore, in contrast to region I, region II exhibits an ESQPT.

In region III, where $\xi_2>0$ and $\delta >2\xi_2$, there are five stationary points. In comparison to region II, the two minima persist at $\bold{r}_2^\pm$, but $\bold{r}_0$  becomes a stable local maximum and two unstable hyperbolic points emerge at $\bold{r}_1^{\pm}$, as seen at the top of Fig.~\ref{fig:clmu2}(d). This implies that $\mathcal{E}_{\bold{r}_2^\pm}<\mathcal{E}_{\bold{r}_1^\pm}<\mathcal{E}_{\bold{r}_0}$. Accordingly, the DOS exhibits an ESQPT$_{\text{peak}}$ at the hyperbolic points energy $\mathcal{E}_{\bold{r}_1^\pm}$ (star) and an ESQPT$_{\text{step}}$ at the local maximum energy $\mathcal{E}_{\bold{r}_0}$ (blue circle). Thus, the transition from region II to region III involves the appearance of an additional ESQPT and no ground-state QPT.

The line of $\xi_{2}=0$ in Fig.~\ref{fig:clmu2}(a), separating region III and region $\widetilde{\mathrm{III}}$, is associated with a first-order phase transition deprived of minima coexistence, similar to the II-$\widetilde{\mathrm{II}}$ transition discussed for $\mu = 1$. For $\delta > 0$, as $\xi_{2} \rightarrow 0$, the four stationary points outside the origin give place to the circular set of global minima shown in Fig.~\ref{fig:nodrive}(a).

\subsection{Three-photon drive}

\begin{figure*}[t]
\centering
\hspace{-0.75cm}
\includegraphics[scale=0.9]{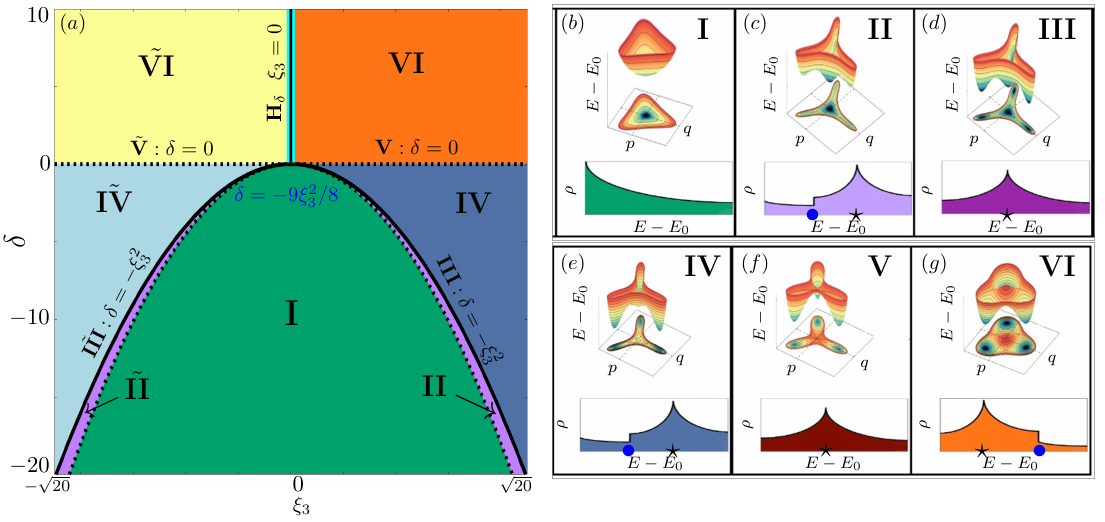} 
\caption{Three-photon KPO. Panel (a): Phase diagram for the detuning parameter $\delta$ and the three-photon drive amplitude $\xi_3$. Panels (b)-(g): energy landscape and phase-space structure (top), and sketches of the density of states (bottom) for the selected cases I-VI, respectively. A ground-state quantum phase transition is represented by a solid line, while excited-state quantum phase transitions are indicated by dotted lines.
}
\label{fig:clmu3}
\end{figure*}

Among the four driving frequencies considered in this work, the three-photon KPO ($\mu=3$) remains the least explored in the literature~\cite{Iachello2023}.
The classical Hamiltonian,
\begin{equation}
\begin{split}
{H}_3^{\mathrm{c}} &= -
\frac{\delta}{2} \left(q^2 + p^2\right)+ \frac{1}{4} \left(q^2 + p^2\right)^2  - \frac{\xi_3}{\sqrt{2}}\left( q^3 - 3qp^2 \right),
\end{split}
\label{qH3}
\end{equation}
has the following seven stationary points and corresponding energies,
\begin{equation}
\begin{array}{ll}
\bold{r}_0 =\{0,0\},
& \hspace{0.4 cm} {\cal E}_{\bold{r}_0}=0,
\\
\\
\bold{r}_1^{\pm} = \left\{Q_\pm, 0\right\},
& \hspace{0.4 cm} {\cal E}_{\bold{r}_1^\pm}={\cal E}_{\pm},
\\ 
\bold{r}_2^{\pm} = \left\{-\dfrac{1}{2}Q_{+}, \pm \dfrac{\sqrt{3}}{2}|Q_+| \right\},
& \hspace{0.4 cm} {\cal E}_{\bold{r}_2^\pm}= {\cal E}_{+},
\\
\bold{r}_3^{\pm}= \left\{-\dfrac{1}{2}Q_{-}, \pm \dfrac{\sqrt{3}}{2}|Q_-| \right\},
& \hspace{0.4 cm} {\cal E}_{\bold{r}_3^\pm}={\cal E}_{-},
\end{array}
\label{sp3}
\end{equation}
where we have defined
\begin{equation}
\begin{array}{ll}
{Q}_{\pm} = \dfrac{1}{2\sqrt2}
\left( 3\xi_3 \pm \sqrt{ 9\xi_3^2 + 8\delta } \right),
&
{\cal E}_\pm =-\dfrac{Q_{\pm}^2}{12}\left(2\delta  + Q_{\pm}^2\right).
\end{array}
\nonumber
\end{equation}
The analysis of the parameters for which these points exist leads to the phase diagram in Fig.~\ref{fig:clmu3}(a) and reveals the phase-space structures shown in Figs.~\ref{fig:clmu3}(b)-(g).

In region I, for $\delta < - 9\xi_3^2/8$, there is a single global minimum at $\bold{r}_0$ and the DOS decreases monotonically as the energy increases [Fig.~\ref{fig:clmu3}(b)]. 

The phase-space structure becomes more complicated at the dotted line $\delta = - 9\xi_3^2/8$ in Fig.~\ref{fig:clmu3}(a), where $Q_+=Q_-=3\xi_3/2\sqrt2$.  
This is a spinodal line,
that marks the appearance of a set of local minima in a first-order phase transition~\cite{Iachello2004}. Hence, in addition to the global minimum at the origin, three degenerate critical points arise according to Eq.~\eqref{sp3}. These new stationary points are inflection points, associated with an ESQPT$_{\text{peak}}$, emerging at positions that respect the rotational symmetry of the system. The point $\bold{r}_1^+=\bold{r}_1^-$ lies exactly on the position axis, and the other two, $\bold{r}_2^+=\bold{r}_3^+$ and $\bold{r}_2^-=\bold{r}_3^-$, can be found by rotating $\bold{r}_1^{\pm}$ at an angle of $2\pi/3$ and $4\pi/3$ around the origin, respectively. 

As the dotted line is crossed and the system enters into region II, characterized by $- 9\xi_3^2/8<\delta<-\xi_3^2$, the three inflection points develop into three local minima and three saddle points, with seven critical points in total. Along the line $p=0$, the critical points at $q=0$, $q=Q_{-}$, and $q=Q_{+}$ respect $0<Q_-<Q_+$ and $\mathcal{E}_{\bold{r}_0}< \mathcal{E}_{+}<\mathcal{E}_{-}$. Over this $q$-axis, the global minimum remains at the origin, $\bold{r}_1^{-} =\{Q_{-},0\}$ is a hyperbolic point, and $\bold{r}_1^{+} =\{Q_+,0\}$ is a local minimum. This structure is repeated by the symmetry operation, and the result is illustrated in Fig.~\ref{fig:clmu3}(c). At the bottom of this figure, the triple local minima and the triple hyperbolic points manifest in the DOS as two ESQPTs: the step-like behavior located at the blue circle and the logarithmic divergence identified by a star, respectively.

The solid line III at $\delta=-\xi_3^2$ indicates a first-order QPT, where now four minima coexist with energy $\mathcal{E}_{\bold{r}_0}= \mathcal{E}_{+}$. In addition, there are three saddle points with energy $\mathcal{E}_{-}$, that are related to the ESQPT$_{\text{peak}}$ illustrated in Fig.~\ref{fig:clmu3}(d). 

The roles of the minima are exchanged when line III is crossed, going from region II to the region IV, where $-\xi_3^2 <\delta < 0$. In region IV, the phase-space origin becomes a local minimum and the points $\bold{r}_1^+$ and $\bold{r}_2^\pm$  are global minima, while the hyperbolic points persist at $\bold{r}_1^{-}$ and $\bold{r}_3^\pm$. Since region~IV has local minima and hyperbolic points, its DOS is equivalent to that of region~II, exhibiting an ESQPT$_{\text{step}}$ and an ESQPT$_{\text{peak}}$, as shown in Fig.~\ref{fig:clmu3}(e).

The dotted line V along $\delta=0$ is an antispinodal line marking a first-order QPT. The local minimum from region IV disappears and the three hyperbolic points $\bold{r}_1^-$ and $\bold{r}_3^\pm$ coalesce at the origin forming a particular type of saddle point, called monkey saddle point. The phase-space topology is now characterized by the monkey saddle point at the origin and the three global minima already present in region IV, as shown in Fig.~\ref{fig:clmu3}(f). Similarly to line III and the dotted line between regions I and II, line V leads only to an ESQPT$_{\text{peak}}$.

The transition from region IV to region VI does not involve a QPT, since in region VI the points $\bold{r}_1^+$ and $\bold{r}_2^\pm$ remain as global minima. However, once the line V is crossed, there are three hyperbolic points $\bold{r}_1^-$ and $\bold{r}_3^\pm$ and, at the  origin, a local maximum, as seen in Fig.~\ref{fig:clmu3}(g). Furthermore, $Q_-$ is now negative, while it is positive in region IV. Therefore, both regions exhibit an ESQPT$_{\text{peak}}$ and an ESQPT$_{\text{step}}$. 

As in the $\mu = 1$ and $\mu=2$ cases, the phase diagram for the three-photon KPO is symmetric for positive and negative values of the control parameter $\xi_{\mu}$ and there is a first-order phase transition line without minima coexistence between regions VI and $\widetilde{\mathrm{VI}}$, which is marked by the solid cyan-highlighted vertical line at $\xi_3 = 0$. 

The classical energy surface analysis is a convenient tool to identify critical points that are associated with QPTs and ESQPTs. However, it is important to keep in mind that in some cases, the detection of ESQPT precursors may require parameters that bring the quantum system significantly close to the classical limit. This is the case, for example, of region II.

\subsection{Four-photon drive}
\label{Sec:Four}

\begin{figure*}[t]
\centering
\includegraphics[scale=0.9]{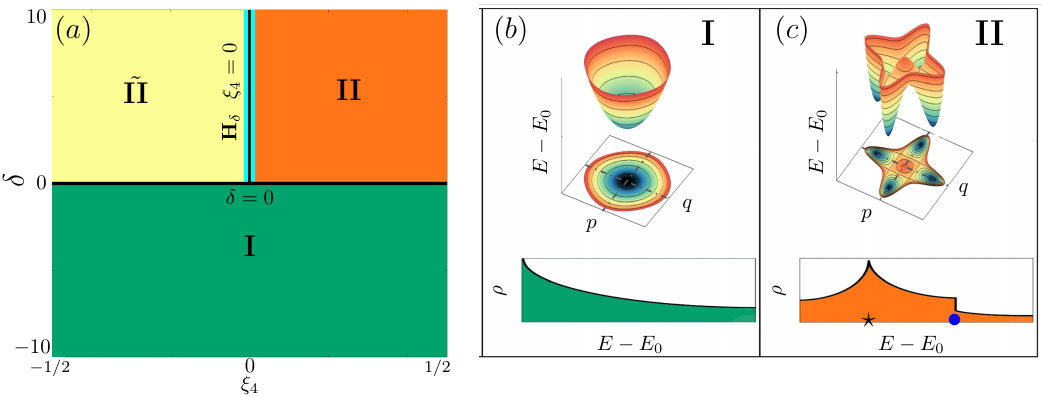} 
\caption{Four-photon KPO. Panel (a): Phase diagram for the detuning parameter $\delta$ and the four-photon drive amplitude $\xi_4$. Panels (b)-(c): energy landscape and phase-space structure (top), and sketches of the density of states (bottom) for phases I-II, respectively. A ground-state quantum phase transition is represented by a solid line.}
\label{fig:clmu4}
\end{figure*}

The four-photon KPO ($\mu=4$) has been considered for error correction schemes~\cite{Kwon2022}. In comparison with the three-photon KPO, its phase diagram is significantly simpler. However, in contrast to the cases with $\mu<4$, the classical Hamiltonian ${H}_4^{\mathrm{c}}$ is unbounded  for  $|\xi_4| \geq 1/2$.  In Sec.~\ref{Sec:less}, we restrict the analysis to $0<\xi_4 < 1/2$ and leave the discussion of the limit $\xi_4 \rightarrow 1/2$ to Sec.~\ref{Sec:more}.

\subsubsection{Analysis for $|\xi_4| < 1/2$}
\label{Sec:less}

The classical Hamiltonian, 
\begin{equation}
\begin{split}
{H}_4^{\mathrm{c}} &= -
\frac{\delta}{2} \left(q^2 + p^2\right)+ \frac{1}{4} \left(q^2 + p^2\right)^2 
- \frac{\xi_4}{2}\left( q^4 - 6q^2p^2 + p^4 \right),
\end{split}
\label{qH4}
\end{equation}
gives the following nine critical points,
\begin{equation}
\begin{array}{lll}
\bold{r}_0=\{0,0\}, 
&& \hspace{0.4 cm} {\cal E}_{\bold{r}_0}=0 ;\\[0.5 ex]
\bold{r}_1^{\pm}= \big\{0, \pm Q_a\big\}, 
&& \hspace{0.4 cm} {\cal E}_{\bold{r}_1^\pm}=-\delta Q_a^2/4  ;\\[0.5 ex]
\bold{r}_2^{\pm}= \big\{\pm Q_a, 0\big\}, 
&& \hspace{0.4 cm} {\cal E}_{\bold{r}_2^\pm}=- \delta Q_a^2/4 ;\\[0.5 ex]
\bold{r}_{3}^{\pm}= \big\{\pm Q_d, \pm Q_d \big\}, 
&& \hspace{0.4 cm} {\cal E}_{\bold{r}_{3}^\pm}=-\delta Q_d^2/2;\\[0.5 ex]
\bold{r}_{4}^{\pm}= \big\{\pm Q_d, \mp Q_d \big\}, 
&& \hspace{0.4 cm} {\cal E}_{\bold{r}_{4}^\pm}=-\delta Q_d^2/2;
\end{array}
\label{sp4}
\end{equation}
where 
$$Q_a=\sqrt{\delta/(1-2\xi_4)} \hspace{0.3 cm} \text{and} \hspace{0.3 cm} Q_d=\sqrt{\delta/[2(1+2\xi_4)]}.$$  
There are only two different phases, I and II, in the diagram of Fig.~\ref{fig:clmu4}(a). 

In region I, where $\delta \leq 0$, there is only one global minimum at $\mathbf{r}_0$, as shown in the top of Fig.~\ref{fig:clmu4}(b). This simple structure reflects in the shape of the DOS seen in the bottom of the same figure. 

The $\delta = 0$ line is a second-order QPT line, where the minimum in the energy landscape at the origin becomes flat (quartic order). Once this line is crossed toward region II, where $\delta > 0$ and $0<\xi_4<1/2$, nine stationary points emerge, as illustrated in Fig.~\ref{fig:clmu4}(c). The origin $\mathbf{r}_0$ now becomes a local maximum, there are four global minima at energy $-\delta Q_{a}^2/4$, and four hyperbolic points at energy $-\delta Q_{d}^2/2$. The DOS in the bottom of Fig.~\ref{fig:clmu4}(c) shows the ESQPT$_{\mathrm{peak}}$ related to the hyperbolic points and the ESQPT$_{\mathrm{step}}$ associated with the local maximum.

\subsubsection{Discussion for  $\xi_4 \rightarrow 1/2$}
\label{Sec:more}

\begin{figure}[t]
\centering
\includegraphics[scale=0.95]{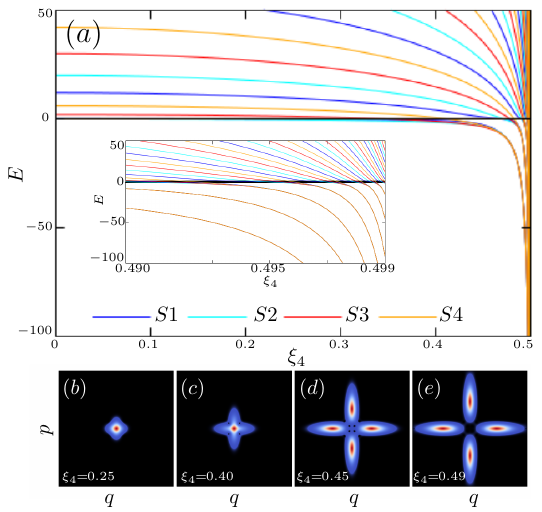} 
\caption{Panel (a): Eigenvalues as a function of the drive amplitude $\xi_4$ for $\delta=0$. Each color represents one of the irreducible sectors: $S1$, $S2$, $S3$, or $S4$. Panels (b)-(e): Husimi function for the ground state at different values of $\xi_4$, approaching $\xi_4\rightarrow 1/2$ from (b) to (e); $N=1200$. 
}
\label{fig:qmu4}
\end{figure}

For $|\xi_4| \ge 1/2$, the classical Hamiltonian becomes associated with unbounded states. This produces important differences between classical and quantum results when the control parameter $|\xi_4| \rightarrow 1/2$. To illustrate this, we show in Fig.~\ref{fig:qmu4}(a) the energy levels for the quantum Hamiltonian with a four-photon drive as a function of $\xi_4>0$ for $\delta=0$. Notice that, contrary to all other figures in this paper, we do not subtract the energy $E$ by the ground-state energy $E_0$ to make it clear that the features not reproduced by the mean-field limit occur in the negative region of the spectrum. When $\xi_4 \rightarrow 1/2$, the classical Hamiltonian tends to
$$
H_4^c \rightarrow -
\frac{\delta}{2} \left(q^2 + p^2\right) + 2 q^2 p^2 ,
$$
which, contrary to the quantum Hamiltonian, can only give positive values when $\delta=0$.

We see that in the region of negative energies in Fig.~\ref{fig:qmu4}(a), the four symmetry sectors (denoted $S1$, $S2$, $S3$, and $S4$)  become quasi-degenerate, as in an effective four-well metapotential [see also the inset in Fig.~\ref{fig:qmu4}(a)]. The appearance of the four branches is supported by the analysis of the Husimi function \cite{Husimi1940}.
\begin{equation}
    Q_{\psi}^{(q,p)} = \frac{1}{\pi} |\langle \alpha (q,p)  | \psi\rangle |^2 ,
    \label{Eq:Husimi}
\end{equation}
for the ground state in Figs.~\ref{fig:qmu4}(b)-(e). 
The Husimi function gives the distribution of a quantum state $|\psi\rangle $ in phase space, where the coherent state $|\alpha\rangle$, from $\hat{a} |\alpha\rangle = \alpha |\alpha\rangle$, defines a point $(q,p)$ in phase space according to $\alpha = (q+ip)/\sqrt{2}$ \cite{Husimi1940}. As $\xi_4$ approaches $1/2$, the Husimi distribution of the ground state changes significantly, from a compact shape centered at the origin of the phase space [Fig.~\ref{fig:qmu4}(b)] to a pattern stretched along the $q$ and $p$ axes [Figs.~\ref{fig:qmu4}(e)] with  probability concentrated (red color) along the axes.

The closing of the energy gap among the four lowest levels, each one from a symmetry sector, extends to higher levels. Whether these changes should be associated with QPTs and ESQPTs will be investigated in a future work, as well as whether they could be captured by modifications of the mean-field limit.

\section{Quantum Tunneling}
\label{Q_tunnel}

\begin{figure*}[t]
\centering
\includegraphics[scale=1.05]{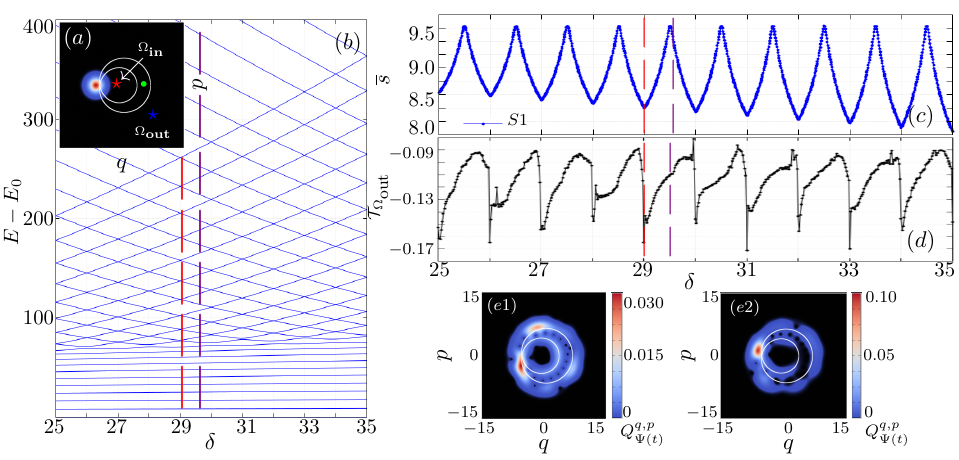}
\caption{Spectral properties and tunneling analysis in the one-photon KPO for $\xi_1=5$. (a) Phase-space structure for $\delta$ in region II of Fig.~\ref{fig:clmu1}: separatrix (white curve), local maximum (red star), and global minimum (green circle). The red circle near the hyperbolic point indicates the initial coherent state considered for the dynamics. Its energy is slightly above the separatrix, being in the phase-space region $\Omega_{\text{out}}$. Trajectories in $\Omega_{\text{out}}$ and $\Omega_{\text{in}}$ can have the same energy. 
(b) Excitation energies as a function of the detuning parameter $\delta$. (c) Average level spacing for the first 150 levels vs $\delta$, showing minima at avoided crossings. (d) Effective tunneling  vs $\delta$ computed for $\Omega_{\text{out}}$ and $Kt \in [0,100]$. A sharp dip in ${\cal T}_{\Omega_{\text{out}}}$ indicates enhanced tunneling away from $\Omega_{\text{out}}$, which happens in the presence of avoided crossings. The red (violet) vertical dashed line in (b)-(d) marks $\delta=29.05$ ($\delta=29.754$). (e1)-(e2) Evolved Husimi function for (e1) $\delta=29.05$, where avoided crossings takes place, and (e2) $\delta=29.754$, where there are no level crossings. Notice the different scales in the color coding for (e1) and (e2), indicating more spreading in (e1). } 
\label{fig:tmu1}
\end{figure*}

In this section, we examine how the spectral level crossings  affect quantum tunneling. Avoided crossings occur within the same symmetry sector and enhance tunneling by enabling state mixing. In contrast, real crossings arise between different symmetry sectors and do not contribute to tunneling. By properly tuning the control parameters $\delta$ and $\xi_{\mu}$, it is therefore possible to enhance or suppress tunneling, which offers a mechanism for controlling quantum dynamics.
 
To quantify quantum tunneling, we employ a measure that captures the spreading of the time-evolved quantum state, $|\Psi(t)\rangle= e^{-i \hat{H}_{\mu} t} |\Psi (0)\rangle$, into specific regions of the phase space $\mathcal{M}$. For this purpose, we use the Husimi function in the position-momentum representation, as defined in Eq.~(\ref{Eq:Husimi}),  where $|\psi\rangle$ is now taken to be $|\Psi(t) \rangle$. To interpret the tunneling dynamics in relation to classical motion, we define and analyze specific invariant regions of the phase space, denoted by $\Omega_k$, which remain unchanged under classical evolution. A region $\Omega_k \subset \mathcal{M}$ is considered invariant if, for any initial condition $x \in \Omega_k$, the corresponding classical trajectory remains confined to $\Omega_k$ for all times $t$, that is, $\varphi_x(t) \in \Omega_k$, where $\varphi: \mathbb{R} \times \mathcal{M} \to \mathcal{M}$ represents the flow generated by the Hamilton's equations of motion given in Eq.~(\ref{Hamil_eq}). 

For a given invariant region $\Omega_k \subset \mathcal{M}$, we define the Husimi volume at time $t$ as~\cite{Villasenor2021},
\begin{equation}
\mathcal{V}_{\Omega_k }\left(t\right) = \iint_{\Omega_k} dqdp \ 
\mathcal{Q}_{\Psi(t)}^{(q,p)},
\end{equation}
which quantifies the fraction of the quantum state's Husimi distribution localized in $\Omega_k $. The Husimi volume is normalized, such that $\mathcal{V} = 1$ when the integral is performed over the entire phase space, i.e. $\Omega_k  = \mathcal{M}$. The computation of the Husimi volume is performed using Monte Carlo integration methods~\cite{Nader2021}.

Finally, we employ the concept of effective tunneling~\cite{Nader2021,Prado2023}, defined as the change in the Husimi volume within a region $\Omega_k$ over a time interval $[t_0, t]$,
\begin{equation}
\mathcal{T}_{\Omega_k}\left(t, t_0\right) = \mathcal{V}_{\Omega_k}\left(t\right) - \mathcal{V}_{\Omega_k}\left(t_0\right).
\label{Eq:Tunneling}
\end{equation}
This measure quantifies the net flow of the quantum state in or out the phase-space region $\Omega_k$ during the specified time interval. For an initial state  concentrated in $\Omega_k$ at $t_0$, a dip in the value $\mathcal{T}_{\Omega_k}$ indicates that the evolved state tunneled out of $\Omega_k$ into other phase-space regions.

\begin{figure*}[t]
\centering
\includegraphics[scale=1.05]{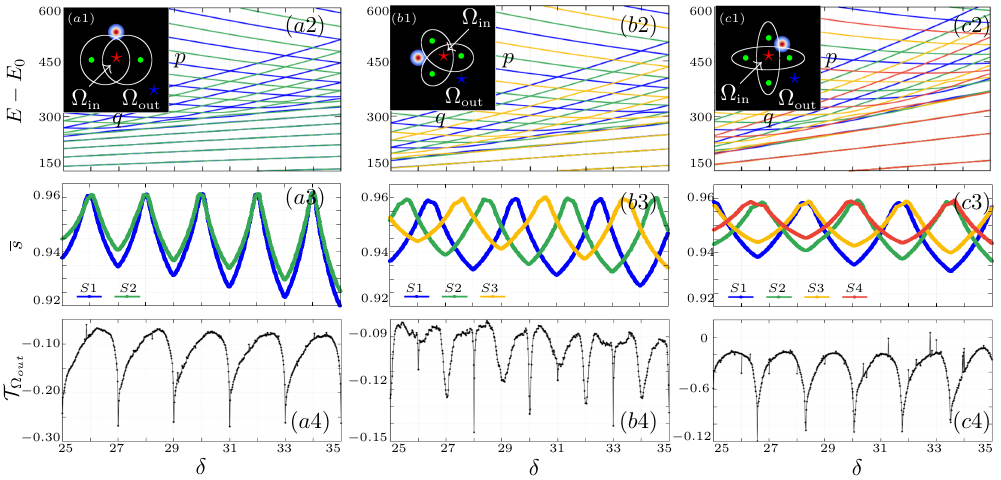}
\caption{Spectral properties and tunneling analysis in (a1)-(a4) the two-photon KPO for $\xi_2=5$, (b1)-(b4) the three-photon KPO for $\xi_3=1$, and (c1)-(c4) the four-photon KPO for $\xi_4=0.25$. (a1)-(c1): Phase-space structure for $\delta$ in (a1) region III of Fig.~\ref{fig:clmu2}, (b1) region VI of Fig.~\ref{fig:clmu3}, and (c1) region II of Fig.~\ref{fig:clmu4}. The figures show the separatrix (white curve),  local maximum (red star), and global minima (green circles). The red circle near the hyperbolic point indicates the initial coherent state considered for the dynamics. Its energy is slightly above the separatrix, being in phase-space region $\Omega_{\text{out}}$. Trajectories in $\Omega_{\text{out}}$ and $\Omega_{\text{in}}$ can have the same energy.
(a2)-(c2): Excitation energies as a function of the detuning parameter $\delta$. Each color indicates a different symmetry sector. Avoided crossings happen in the same symmetry sector. (a3)-(c3): Average level spacing for each symmetry sector $Sk$ vs $\delta$, showing minima at avoided crossings. Each curve considers the first 150 levels.  (a4)-(c4): Effective tunneling vs $\delta$ computed for $\Omega_{\text{out}}$ and $Kt \in [0,100]$. A sharp dip in ${\cal T}_{\Omega_{\text{out}}}$ indicates enhanced tunneling away from $\Omega_{\text{out}}$, which happens in the presence of avoided crossings.
}
\label{fig:tmu234}
\end{figure*}

\subsection{Tunneling in the absence of symmetry sectors}

We begin the analysis with the one-photon ($\mu=1$) KPO, where there are no symmetry sectors, so any crossing is necessarily avoided. We focus on phase II of Fig.~\ref{fig:clmu1}. As shown in Fig.~\ref{fig:tmu1}(a) [see also Fig.~\ref{fig:clmu1}(c)], the classical phase space features a global minimum (green circle), a local maximum (red star), and a hyperbolic point at the crossing of separatrix (white line). The figure also indicates the phase-space region $\Omega_{\text{out}}$, with energies above the separatrix, and region $\Omega_{\text{in}}$, with energies between the separatrix and the local maximum. Trajectories within these regions may have the same energy. When this happens, there are avoided energy crossings in the quantum spectrum.

In Fig.~\ref{fig:tmu1}(b), we show the spectrum of the Hamiltonian in Eq.~(\ref{qH1}) as a function of the detuning parameter $\delta$, fixing $\xi_1=5$. For intermediate energies, between the ESQPT$_{\text{peak}}$ energy (approximately the energy of the hyperbolic point) and the ESQPT$_{\text{step}}$ energy (approximately the energy of the local maximum), the spectrum exhibits avoided crossings for integer values of the detuning parameter. These values can be obtained using a semiclassical approach
based on the Einstein-Brillouin-Keller quantization rule~\cite{Prado2023} that determines for which $\delta$, pairs of classical trajectories have the same energy.
    
In Fig.~\ref{fig:tmu1}(c), we calculate the average level spacing for the same parameters used in Fig.~\ref{fig:tmu1}(b). The average is performed over the spectral region that goes from the ground state energy to energies above the step-like ESQPT, so that the entire energy region of level crossings is included. As expected, the minima in the average level spacing align with the energy values at which avoided crossings occur. 

To probe quantum tunneling, in Fig.~\ref{fig:tmu1}(d),  we compute the effective tunneling measure defined in Eq.~(\ref{Eq:Tunneling}) for an initial coherent state centered slightly above the hyperbolic point [see red circle in Fig.~\ref{fig:tmu1}(a)]. This places the initial state's energy within the phase-space region $\Omega_{\text{out}}$, so we compute the effective tunneling with respect to $\Omega_\text{out}$. The observed dips in the values of $\mathcal{T}_{\Omega_\text{out} }$ correlate with the minima in the average level spacing [Fig.~\ref{fig:tmu1}(c)] and indicate enhanced tunneling away from $\Omega_\text{out} $ caused by avoided crossings. 

In Figs.~\ref{fig:tmu1}(e1)-(e2), we display the Husimi function of the evolved state at time $t=100/K$. The detuning parameter corresponding to Fig.~\ref{fig:tmu1}(e1) is marked by a vertical red dashed line in Figs.~\ref{fig:tmu1}(b)–(d), while the value used in Fig.~\ref{fig:tmu1}(e2) is indicated by a vertical violet dashed line in the same panels. In Fig.~\ref{fig:tmu1}(e1), where  the detuning parameter leads to avoided crossings, we observe tunneling from the phase-space region $\Omega_{\text{out}}$ to $\Omega_{\text{in}}$. In contrast,  Fig.~\ref{fig:tmu1}(e2) shows no spread into region $\Omega_{\text{in}}$, consistent with the suppression of tunneling due to the absence of avoided crossings.

\subsection{Tunneling in the presence of symmetries sectors}

We now examine tunneling in $\mu$-photon KPOs with $\mu=2,3,4$, in which cases the system decomposes into $2, 3$, and $4$ irreducible sectors, respectively, due to its underlying $\mathbb{Z}_{\mu}$  symmetry. For $\mu=2$, we focus on phase III of Fig.~\ref{fig:clmu2}, for $\mu=3$ on phase VI of Fig.~\ref{fig:clmu3}, and for $\mu=4$ on phase II of Fig.~\ref{fig:clmu4}. As in the previous subsection, the classical phase space for all three cases shown in Figs.~\ref{fig:tmu234}(a1)-(c1) reveal a common structure: an outer phase-space region $\Omega_{\text{out}}$ with energy above the separatrix (white line) and an inner region $\Omega_{\text{in}}$ bounded by the separatrix energy and a local maximum (red star). Depending on the control parameters, classical trajectories in $\Omega_{\text{in}}$ and $\Omega_{\text{out}}$ can have the same energy, which manifests in the quantum regime as real or avoided level crossings in the spectrum.

We investigate tunneling between $\Omega_{\text{out}}$ and $\Omega_{\text{in}}$. Notice that for $\mu=2,3,4$, there are other phase-space regions associated with the global minima (green circles) seen in Figs.~\ref{fig:tmu234}(a1)-(c1). These regions have the same energy and could be considered for studies of tunneling as well.

The spectrum of the quantum Hamiltonian $\hat{H}_{\mu}$ in Eq.~(\ref{Eqeff}) is shown as a function of the detuning parameter $\delta$ in Figs.~\ref{fig:tmu234}(a2), (b2), and (c2) for $\mu=2$, $3$, and $4$, respectively. Each color represents a symmetry sector. Below the ESQPT$_{\text{peak}}$, the spectrum is $\mu$-fold degenerate. Between the ESQPT$_{\text{peak}}$ and the ESQPT$_{\text{step}}$, energy crossings can occur depending on the control parameter values. Real level crossings happen between different symmetry sectors and avoided crossings within levels in the same symmetry sector. Above the ESQPT$_{\text{step}}$ energy, the spectrum does not exhibit level crossings. 

For the two-photon KPO, all crossings are real for even values of the detuning parameter and  avoided for odd values of $\delta$ \cite{Prado2023,Iachello2023,Venkatraman2024}. Unlike this case, for three- and four-photon KPOs, real and avoided crossings can coexist for the same value of $\delta$, resulting in a more intricate spectral structure. In Figs.~\ref{fig:tmu234}(a3), (b3), (c3), we show the average level spacing, $\overline{s}$, for each symmetry sector {\em separately} as a function of the detuning parameter $\delta$ for $\mu=2,3,4$, respectively. Small values of $\overline{s}$ (avoided crossing) for one symmetry may coincide with large values of $\overline{s}$ (real crossing) for another symmetry only when $\mu=3,4$, as seen in Figs.~\ref{fig:tmu234}(b3) and (c3).

For the two-photon KPO in Fig.~\ref{fig:tmu234}(a3), the values of $\overline{s}$  are correlated for both symmetries, with minimum values corresponding to odd $\delta$ and maximum values to even $\delta$. 
For the three-photon KPO in Fig.~\ref{fig:tmu234}(b3), the lowest values of $\overline{s}$ alternate among the three symmetry sectors, and we have not identified any simple function for their dependence on $\delta$. For the four-photon KPO in Fig.~\ref{fig:tmu234}(c3), a pattern emerges where the lowest values of the average level spacings appear simultaneously for two symmetry sectors. There are now two pairs of correlated symmetry sectors, a pair with even parity and the other with odd parity. Whenever two sectors exhibit avoided crossings for a value of $\delta$, the other two show real crossings.  As in the three-photon case, no straightforward rule dictates the dependence of the lowest values of $\overline{s}$ on $\delta$.

In Figs.~\ref{fig:tmu234}(a4), (b4), (c4), we finally compute the effective tunneling
measure defined in Eq.~(\ref{Eq:Tunneling}) for an initial coherent state centered slightly above the hyperbolic point [see red circle in Figs.~\ref{fig:tmu234}(a1)-(c1)]. This initial state has energy within the phase-space region $\Omega_{\text{out}}$, so we compute $\mathcal{T}_{\Omega_\text{out} }$. In all three cases, $\mu=2,3,4$, tunneling enhancement in Figs.~\ref{fig:tmu234}(a4)-(c4) correlates with the lowest values of $\overline{s}$ in Figs.~\ref{fig:tmu234}(a3)-(c3). Therefore, for the two-photon KPO, tunneling amplification away from $\Omega_\text{out}$ towards $\Omega_\text{in}$ happens when both symmetry sectors exhibit avoided crossings, while for the three- and four-photon KPO, this happens when at least one sector has avoided crossings.

\section{CONCLUSION}
\label{Conclusion}

We provided a systematic analysis of ground-state and excited-state quantum phase transitions (QPTs and ESQPTs) in Kerr parametric oscillators (KPOs) subject to one-, two-, three-, or four-photon drive. By deriving and analyzing the corresponding classical Hamiltonians, we identified critical points associated with first- and second-order QPTs and ESQPTs. The results reveal rich phase-diagram structures influenced by the control parameters and discrete $\mathbb{Z}_{\mu}$ symmetries.

The $\mathbb{Z}_{\mu}$  symmetry partitions the Hilbert space into symmetry sectors and affects the nature of spectral crossings. Real level crossings occur between different symmetry sectors, while avoided crossings occur within the same sector. This difference directly impacts quantum tunneling dynamics: avoided crossings enable tunneling via state hybridization, whereas real crossings inhibit tunneling by protecting states from mixing. Therefore, quantum tunneling can be selectively enhanced or suppressed by tuning the system parameters, thus offering a mechanism for dynamical control in KPOs.

In KPOs with three- or four-photon drive, real and avoided crossings may coexist for different symmetry subspaces. We verified that significant tunneling can occur even when avoided crossings is confined to a single symmetry sector. 

In the four-photon case, novel behavior emerges as the system approaches the unbounded Hamiltonian regime ($\xi_4 \rightarrow 1/2$). In this limit, the quantum system undergoes changes that are not captured by the classical mean-field approximation. The analysis of the Husimi function reveals the onset of potentially unexplored QPT and ESQPT phenomena, that calls for further investigation. 

Overall, our study offers a unified framework for understanding the interplay between symmetry, quantum phase transitions, and tunneling in driven nonlinear quantum oscillators. We demonstrated that by adjusting the detuning and drive amplitude, one can engineer the spectrum, which can have direct implications for critical quantum sensing and dynamical control, particularly in experimental platforms that realize KPOs, such as superconducting circuits and optical devices.

\begin{acknowledgments}
The authors wish to thank Francesco Iachello, Octavio Casta\~nos, and Manuel Calixto for insightful discussions and suggestions that have helped in shaping this paper. The authors acknowledge support from the National Science
Foundation Center for Quantum Dynamics on Modular Quantum Devices (CQD-MQD) under Award Number
2124511. This project has also received funding from the Grant PID2022-136228NB-C21 funded by MICIU/AEI/
10.13039/501100011033 and, as appropriate, by ``ERDF A way of making Europe'', by ``ERDF/EU'', by the ``European Union'', or by the ``European Union NextGenerationEU/PRTR'' and by the FEDER-UHU project POSH-AI, EPIT1462023. Computing resources supporting this work were partly provided by the CEAFMC and Universidad de Huelva High Performance Computer (HPC@UHU) located in the Campus Universitario el Carmen and funded by FEDER/MINECO project UNHU-15CE-2848.
\end{acknowledgments}

\appendix

\section{Classical limit}
\label{Ap:CH}

To derive the classical Hamiltonian and obtain a continuous spectrum, we introduce the parameter $N_{\text{eff}}$ in
\begin{equation}
\hat{a}=\sqrt{\frac{N_{\text{eff}}}{2}}(\hat{q}+i\hat{p})
\quad\mathrm{and}\quad
\left[ \hat{q},\hat{p} \right]=\frac{i}{N_{\text{eff}}}.
\label{ap1}
\end{equation}
From the above definitions, we then have 
$$ \hat{q} = \frac{\hat{a}^{\dagger} + \hat{a}}{\sqrt{2N_{\text{eff}}}} \quad\mathrm{and}\quad 
\hat{p} = \frac{i(\hat{a}^{\dagger} - \hat{a})}{\sqrt{2 N_{\text{eff}}}} ,
$$
showing that  $1/\sqrt{ N_{\text{eff}} }$ corresponds to the zero-point fluctuation of a harmonic oscillator. The classical limit is achieved by taking $N_{\text{eff}}\rightarrow\infty$.

Applying Eq.~\eqref{ap1} to the quantum Hamiltonian in Eq.~\eqref{Eqeff} leads to
\begin{equation}
\begin{split}
\frac{\hat{H}_\mu}{K}  = &
-\frac{\delta N_{\text{eff}} }{2}\left(\hat{q}-i\hat{p}\right)
\left(\hat{q}+i\hat{p}\right)
\\&+ 
\frac{N_{\text{eff}}^2}{4}\left(\hat{q}-i\hat{p}\right)^2
\left(\hat{q}+i\hat{p}\right)^2
\\&-
\frac{\xi_\mu N_{\text{eff}}^{\mu/2}}{2^{\mu/2}}
\left[\left(\hat{q}-i\hat{p}\right)^\mu+
\left(\hat{q}+i\hat{p}\right)^\mu\right].
\end{split}
\end{equation}
In the limit $N_{\text{eff}}\rightarrow\infty$, the quantum operators approach classical coordinates, $\hat{q}\rightarrow q$ and $\hat{p}\rightarrow p$, resulting in the classical Hamiltonian $H_\mu^c$. We find that
\begin{equation}
\frac{H_\mu^c}{ K^c }=-\frac{\delta^c}{2}\left(q^2 + p^2\right) +\frac{1}{4}\left(q^2 + p^2\right)^2
- \frac{2\xi_\mu^c}{2^{\mu/2}} \mathcal{F}_\mu,
\end{equation}
where the classical parameters can be mapped back to their counterparts in the quantum Hamiltonian through the equations for the Kerr nonlinearity, $K^c=KN_{\text{eff}}^2$, for the detuning $\delta^c=\delta/N_{\text{eff}}$, and for  $\xi_\mu^c=\xi_\mu / N_{\text{eff}}^{2-\mu/2}$. The function $\mathcal{F}_\mu$ is defined as $\Re\left[(q+ip)^\mu\right]$. In the main text, we conveniently fix $N_{\text{eff}}=1$.

\section{Phase transition order}
\label{Ap:Op}

\begin{figure}[t]
\centering
\includegraphics[scale=0.335]{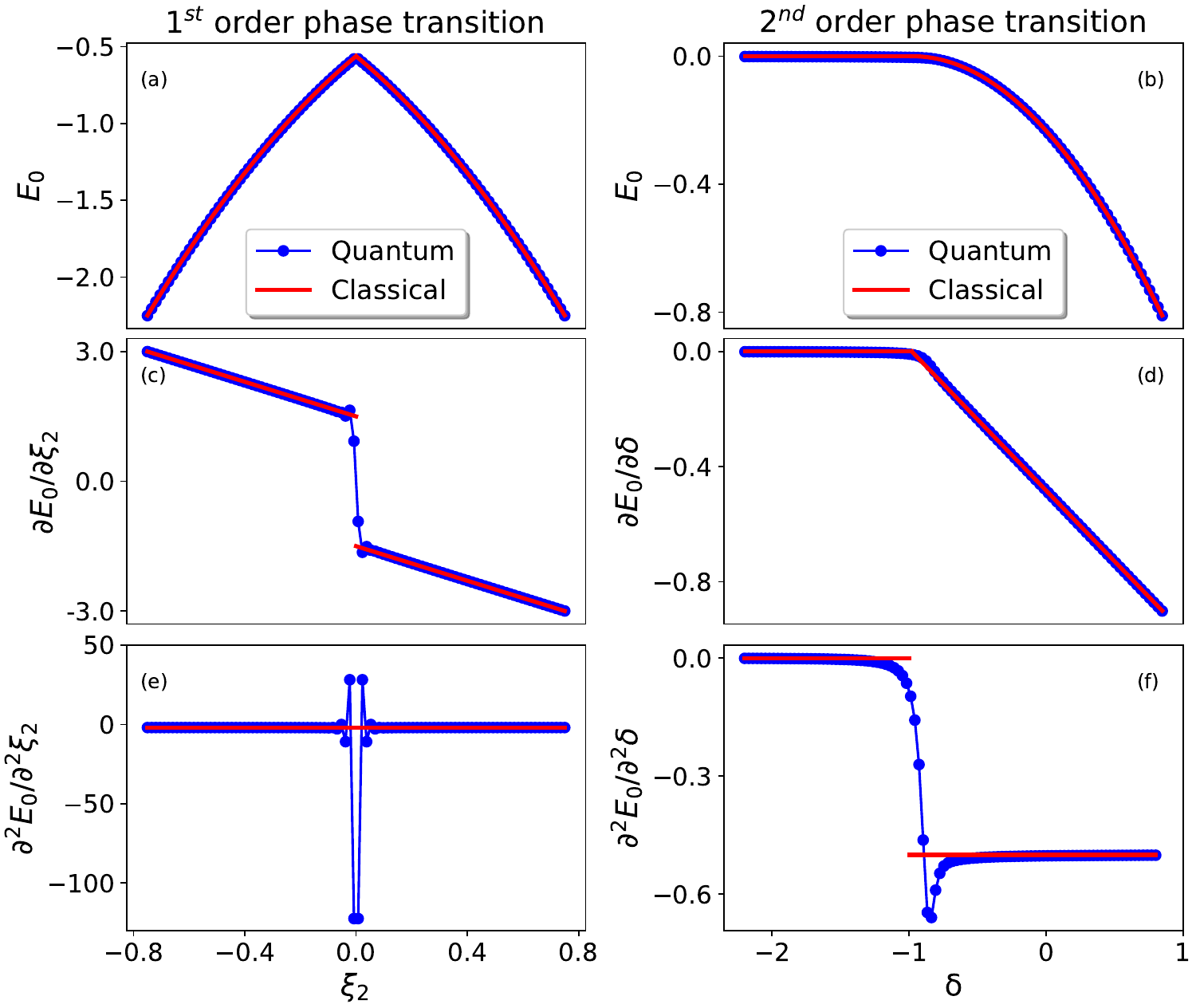} 
\caption{Analysis of the order of the phase transition for the case $\mu=2$.  
Left panels illustrate the transition between regions III and $\widetilde{\mathrm{III}}$ for $\delta=1.5$. Panels (a), (c), and (e) present, respectively, the lowest energy $E_0$, its first- and second-order derivatives as a function of $\xi_2$. 
Right panels refer to the transition between regions I and II for $\xi_2=0.5$. Panels (b), (d), and (f) present, respectively, the lowest energy $E_0$, its first- and second-order derivatives as a function of $\delta$. 
The blue circles (red lines) refer to the quantum (classical) results. 
All panels adopt $N = 4000$ and $N_\text{eff} = 100$, ensuring that quantum and classical results are very close. 
}
\label{fig:order_tran}
\end{figure}

In our analysis, we use Ehrenfest's criterion for the lowest energy $E_0$ and its derivatives to classify the QPTs. This is illustrated in Fig.~\ref{fig:order_tran} for the case $\mu=2$, where the left panels [Figs.~\ref{fig:order_tran}(a), (c), and (e)] correspond to the transition between regions III and $\widetilde{\mathrm{III}}$ and the right panels [Figs.~\ref{fig:order_tran}(b), (d), and (f)] refer to the transition between regions I to II. Blue points represent the quantum results and red lines the classical values.

Let us begin by discussing the left panels. We fix $\delta=1.5$ and study $E_0$ [Fig.~\ref{fig:order_tran}(a)], its first-order derivative $\partial E_0/\partial \xi_2$ [Fig.~\ref{fig:order_tran}(c)], and its second-order derivative $\partial^2 E_0/\partial \xi_2^2$ [Fig.~\ref{fig:order_tran}(e)] as a function $\xi_2$. The discontinuity of the first-order derivative in Fig.~\ref{fig:order_tran}(c) indicates a first-order phase transition.

Let us now focus on the right panels. We fix $\xi_2=0.5$ and study $E_0$ [Fig.~\ref{fig:order_tran}(b)], its first-order derivative $\partial E_0/\partial \delta$ [Fig.~\ref{fig:order_tran}(d)], and its second-order derivative $\partial^2 E_0/\partial \delta^2$ [Fig.~\ref{fig:order_tran}(f)] as a function $\delta$. The discontinuity of the second-order derivative in Fig.~\ref{fig:order_tran}(f) indicates a second-order phase transition.

This methodology can be systematically applied to any QPT across different cases.



%

\end{document}